\documentclass[twocolumn,showpacs,preprintnumbers,amsmath,amssymb]{revtex4}

\usepackage{graphicx}
\usepackage{dcolumn}
\usepackage{bm}
\usepackage{latexsym}
\usepackage{amsfonts}
\usepackage{amssymb}
\usepackage{amsmath}

\begin{document}
\preprint{KUNS-1986}
\title{Moduli Stabilization in String Gas Compactification}
\author{Sugumi Kanno}
\email{sugumi@tap.scphys.kyoto-u.ac.jp}
\author{Jiro Soda}
\email{jiro@tap.scphys.kyoto-u.ac.jp}
\affiliation{
 Department of Physics, Kyoto University, Kyoto 606-8501, Japan
}%

\date{\today}

\begin{abstract}
We investigate the moduli stabilization in string gas compactification.
 We first present a numerical evidence showing the
 stability of the radion and the dilaton. To understand this numerical result,
 we construct the 4-dimensional effective action by taking 
 into account T-duality. It turns out that the dilaton is actually
 marginally stable.  When the moduli other than the dilaton is 
 stabilized at the self-dual point, 
 the potential for the dilaton disappears and then
 the dilaton is stabilized due to the hubble damping. 
 In order to investigate if this mechanism works in more general cases,
 we analyze the stability of $T_2 \otimes T_2 \otimes T_2$ compactification
 in the context of massless string gas cosmology.
 We found that the volume moduli, the shape moduli, and the flux moduli 
 are stabilized at the self dual point in the moduli space.
 Thus, it is proved that this simple compactification model is stable.
\end{abstract}

\maketitle

\section{Introduction}

It is widely believed that the superstring theory is the most promising
 candidate for the quantum theory of gravity. 
  The most attractive feature of the superstring theory is the existence of 
  the target space duality ( T-duality )~\cite{Kikkawa:1984cp}. 
 It is T-duality that implies the minimal length scale, i.e. 
 the string length scale $\ell_s$ (we take the unit $\ell_s =1$ 
 throughout this paper). Thus, there is a possibility 
 to avoid the cosmological initial singularity in the superstring theory. 
 Another feature of the superstring theory is the presence of extra-dimensions.
 Therefore, it is inevitable to study the higher-dimensional cosmology
 and explain how 4-dimensional large external space emerges. 
 Brandenberger and Vafa proposed an interesting cosmological 
 scenario~\cite{Brandenberger:1988aj,Tseytlin:1991xk}
(see also previous works 
\cite{Nishimura:1986wp,Matsuo:1986es,Kripfganz:1987rh}).
 They argued the avoidance of the cosmological singularity 
 due to T-duality and proposed a mechanism how only 
  3 spatial dimensions become large through 
  the annihilation of  winding modes  
(see also \cite{Easther:2002mi,Easther:2004sd} 
concerning this point). 
Recent developments of D-brane physics stimulates the 
study of string gas or brane 
gas scenario~\cite{Alexander:2000xv,Easson:2001fy,Boehm:2002bm,Watson:2002nx,Campos:2003gj,Bassett:2003ck,Easson:2005ug,Kaya:2005qm}.

Because this idea is so attractive, it is important to clarify 
 the issue of moduli stabilization in this scenario.
The purpose of this paper is to reveal
 to what extent the moduli can be stabilized
in the scenario of Brandenberger and Vafa. 

Historically, Watson and Brandenberger first
 demonstrated the stability of the radion but in their work
the dilaton runs logarithmically~\cite{Watson:2003gf,Berndsen:2004tj}.
The 4-dimensional effective action is also obtained 
and concluded neither the dilaton nor the radion can be stabilized except 
for 5-dimensional case~\cite{Battefeld:2004xw}.
The effects of inhomogeneous perturbations are investigated and it is shown 
that they do not affect the stability of the radion~\cite{Watson:2003uw}.
The importance of the massless string modes are recently
recognized~\cite{Watson:2004aq,Patil:2004zp,Patil:2005fi}.
The effects of D-string gas is also studied~\cite{Patil:2005nm}.
However, all of the previous analysis are restricted to the special subspace
of the moduli space and the role of T-duality has not been fully explored. 
In this paper, we investigate moduli stabilization in the string gas
compactification generally with focusing on  T-duality.

The organization of this paper is as follows.
In sec.II,  we review  T-duality in the low energy effective action 
of string theory. We also present a string gas model 
as the T-duality invariant matter. 
In sec.III, we present the numerical calculations of the simplest case 
which show the stability of the radion and the dilaton. 
In sec.IV, we obtain the T-duality invariant 
4-dimensional effective action and clarify
 why the dilaton is stabilized in our numerical results.
 In sec.V, using the 4-dimensional effective action, 
 we show the stability of $T_2 \otimes T_2 \otimes T_2$ compactification. 
 The final section is devoted to the conclusion.

\section{T-duality in Cosmology}

Here, we would like to review T-duality in string theory
with focusing on its relation to cosmology.
In the low energy effective action of string theory,
there exists the $O(6,6,{\bf R})$ symmetry which includes the T-duality
symmetry $O(6,6,{\bf Z})$
as a special case~\cite{Gasperini:1991ak,Giveon:1994fu,Lidsey:1999mc,Gasperini:2002bn}.
 In the full string theory, 
$O(6,6,{\bf R})$ symmetry cease to exist. However, the T-duality symmetry
$O(6,6,{\bf Z})$ remains. 
 In fact, in the case of a string propagating
in constant background fields, the T-duality symmetry $O(6,6,{\bf Z})$
 exists in the mass spectrum of a quantum string. 
In the cosmological background, we do not know exact spectrum.
Here, we treat the gas of strings as test objects and take the metric
in a self-consistent manner. It is usual to do so in cosmology.

\subsection{T-duality in Low Energy Effective Action}

The bosonic part of the low energy effective action of 
the superstring theory takes the following form
\begin{eqnarray}
S = \frac{1}{2\kappa^2 }\int d^{10} x \sqrt{-G} e^{-2\phi}
	\left[ 
	\overset{(10)}{R} + 4 (\nabla \phi )^2 - \frac{1}{12} H^2 
	\right]  \ , 
\end{eqnarray}
where $G_{AB}$ and $\phi$ denote the 10-dimensional metric 
with $A,B = 0, 1, \cdots , 9$ and the dilaton, respectively.
Here, we used the notation $(\nabla \phi)^2 = \partial^A \phi \partial_A \phi $
and $H= dB$ is the field strength of 
the anti-symmetric tensor field $B_{AB}$. We also defined the 10-dimensional
 gravitational coupling constant $\kappa$.  

We assume the 4-dimensions are selected by the Brandenberger-Vafa
mechanism. Hence, 
we consider the cosmological ansatz for the metric:
\begin{eqnarray}
ds^2 = g_{\mu\nu} (x^\mu ) dx^\mu dx^\nu + \gamma_{ab} (x^\mu) dy^a dy^b  \ ,
\end{eqnarray}
where $g_{\mu\nu}$ is the metric of 4-dimensional external spacetime and
$\gamma_{ab}$ is the metric of the internal 6-dimensional compact space. Here, 
both metric are assumed to depend only on 4-dimensional coordinates $x^\mu$.
 This means the internal space is flat with respect to  $y^a$ . 
It is convenient to define shifted dilaton $\bar{\phi}$ by
\begin{eqnarray}
  \sqrt{\gamma} e^{-2\phi} \equiv e^{-2\bar{\phi}} \ .
\end{eqnarray}
Now, we define the 6$\times$6 matrix $(\Gamma)_{ab} =\gamma_{ab}$
in terms of the internal space components of the metric. We assume the
anti-symmetric field $B_{AB}$ exists only in the internal space defined
by the 6$\times$6 matrix, $(B)_{ab}=B_{ab}$  depending
only on 4-dimensional coordinate $x^\mu$. Then the action can be set
into a more compact form by using the 12 $\times$ 12 matrix $Q$: 
\begin{eqnarray}
Q = \left(
\begin{array}{ll}
\Gamma^{-1}  & \quad -\Gamma^{-1} B \\
B\Gamma^{-1} & \quad \Gamma - B \Gamma^{-1} B 
\end{array}
\right)
\end{eqnarray}
which satisfies a symmetric matrix element of the pseudo-
orthogonal $O(6,6,{\bf R})$ group, since
\begin{eqnarray}
    Q^{T} \eta Q = \eta \ , \quad Q^{T} = Q   \ ,
\end{eqnarray}
for any $B$ and $\Gamma$. Here, $\eta$ consists of
the unit 6-dimensional matrix $I$,
\begin{eqnarray}
 \eta = \left(
 \begin{array}{cc}
 0 & I \\
 I & 0
 \end{array}
 \right)  \ .
\end{eqnarray}
Using the metric (2) and the variables (3) and (4), 
the action can be written as
\begin{eqnarray}
 S= \frac{V_6}{2\kappa^2} \int d^{4} x e^{-2\bar{\phi}}
 \left[ R + 4 (\partial \bar{\phi})^2 
 + \frac{1}{8} {\rm Tr} \partial^\mu {Q} \partial_\mu Q^{-1}   \right] \ ,
\end{eqnarray}
where $V_6$ is the coordinate volume of the internal space
and $R$ is the 4-dimensional scalar curvature. 
Here, $(\partial \bar{\phi})^2$ represents 
$ \partial^\mu \bar{\phi} \partial_\mu \bar{\phi}$ and ${\rm Tr}$ denotes
the trace of the matrix. One can see
the action is invariant under $O(6,6, {\bf R})$ transformation 
\begin{eqnarray}
Q\rightarrow\tilde{Q} = \Lambda^T Q \Lambda \ ,
\label{odd}
\end{eqnarray}
where $\Lambda$ is the $12 \times 12$ matrix satisfying 
$\Lambda^T \eta \Lambda =\eta$.
Note that the shifted dilaton is invariant under this $O(6,6,{\bf R})$ 
transformation $\bar{\phi}\rightarrow\bar{\phi}$.
The special $O(6,6, {\bf R})$ transformation represented by $\Lambda =\eta$ 
belongs to T-duality transformation.  
More explicitly Eq.~(\ref{odd}) gives, 
\begin{eqnarray}
\tilde{\Gamma} 
	&=& \left( \Gamma - B\gamma^{-1} B \right)^{-1}
	\label{Gamma}\\
\tilde{B} 
	&=& -\Gamma^{-1} B \left( \Gamma - B\Gamma^{-1} B \right)^{-1}
	\label{B}
\end{eqnarray}
When we set $B=0$, this corresponds to an inversion of the internal space 
matrix, $\tilde{\Gamma} = \Gamma^{-1}$.
So far, we have seen only the kinetic part.
It is interesting to see if the potential energy for the moduli
can be induced by the string gas. 
If yes, because of T-duality, one can expect the moduli in the 
internal space are stabilized at the self-dual point, $\tilde{\Gamma}=\Gamma$
 and $\tilde{B} =B$.
This is the subject of the next subsection.

\subsection{T-duality Invariant String Gas}

Let us consider a closed string in the constant background field
$g_{\mu\nu} , \gamma_{ab} , B_{ab}$. 
The action for the string with the position $X^A$
 is given by the nonlinear sigma model,
\begin{eqnarray}
S&=&-\frac{1}{4\pi}\int d\sigma d\tau
	\left[
	G_{AB}\partial^{\cal M} X^A \partial_{\cal M} X^B
	    \right. \nonumber \\
  && \qquad \qquad \qquad \qquad \quad \left. 	
  +\epsilon^{\cal MN}B_{AB}\partial_{\cal M}
  	X^A \partial_{\cal N}X^B
  	\right] \nonumber\\
  	&\equiv&-\frac{1}{4\pi\alpha'}\int d\sigma d\tau{\cal L}
	\label{sg}
\end{eqnarray}
where indices $\{\cal{M,N, \cdots}\}$ are used for tensors on
a 2-dimensional world-sheet which can be described
in terms of two parameters $ X^A (\tau , \sigma )$. 
Defining variables
\begin{eqnarray}
P^\tau_A&\equiv&\frac{\partial\cal L}{\partial \dot{X}^A}=
	\frac{1}{2\pi } \left[
	G_{AB} \dot{X}^B + B_{AB} X^{\prime B} \right] \ , \\
	P^\sigma_A 
	&\equiv&\frac{\partial\cal L}{\partial X^{\prime A}}=
	-\frac{1}{2\pi} \left[
	G_{AB} X^{\prime B} + B_{AB} \dot{X}^{ B} \right]  \ , 
\end{eqnarray}
where a dot and a prime denote a $\tau$- and a $\sigma$-derivative,
respectively.
Here,  we should keep it in mind that each component is the following:
$G_{\mu\nu} = g_{\mu\nu} , G_{ab}= \gamma_{ab} , G_{\mu a} =0 $
and $B_{\mu\nu} = B_{\mu a} =0 , B_{ab} \neq 0 $. 
The variation of the action (11) yields the equations of motion
\begin{eqnarray}
    \dot{P}^\tau_A+P^{\prime\sigma}_A=0 \ .
\end{eqnarray}
This can be simplified to
\begin{eqnarray}
  \ddot{X}^A - X^{\prime\prime A} =0 \ .
\end{eqnarray}
Notice that $B_{ab}$ does not appear in the equation of motion of 
a string. This is because $B_{ab}$ becomes a total derivative in (\ref{sg}).
In the case of a closed string, the general solution can be written as a sum 
of the left-moving and the right-moving solutions:
\begin{eqnarray}
  X^A (\tau , \sigma ) = X^A_L (\tau +\sigma) + X^A_R (\tau -\sigma )
\end{eqnarray}
where 
\begin{eqnarray}
X^A_L (\tau +\sigma)&=& \frac{1}{2} x^A_L 
	+ \frac{1}{\sqrt{2}}\bar{\alpha}^A_0 (\tau + \sigma ) \nonumber\\
	&&\qquad + i \frac{1}{\sqrt{2}} \sum_{n\neq 0} 
	\frac{\bar{\alpha}^A_n}{n}  e^{-in(\tau +\sigma )}  
	\label{left}
\end{eqnarray}
and
\begin{eqnarray}	
 X^A_R (\tau -\sigma)&=& \frac{1}{2} x^A_R 
	+ \frac{1}{\sqrt{2}}  \alpha^A_0 (\tau -\sigma ) \nonumber\\
	&&\qquad  + i \frac{1}{\sqrt{2}} \sum_{n\neq 0} 
	\frac{\alpha^A_n}{n}  e^{-in(\tau -\sigma )}  \ .
	\label{right}
\end{eqnarray}
Here,  $x^A_L , x^A_R , \alpha^A_n , \bar{\alpha}^A_n$
are the expansion coefficients which become the operators when quantized.

The momentum of the center of mass is given by
\begin{eqnarray}
p_A &\equiv& \int^{2\pi}_0 d\sigma P^\tau_A \nonumber\\
	&=& \frac{1}{\sqrt{2}}
	\left[ G_{AB} (\bar{\alpha}^B_0 + \alpha^B_0 ) 
	+ B_{AB} (\bar{\alpha}^B_0 -\alpha^B_0 ) \right] \ ,
	\label{momentum}
\end{eqnarray}
This is a conserved quantity $\dot{p}_A=0$.
For the compact internal dimensions, 
$p_a$ is quantized to be an integer.

A closed string may wind around the compact direction. The winding 
$w^a$ boundary condition 
$X^a (\tau ,\sigma + 2\pi) = X^a (\tau ,\sigma)+ 2\pi w^a $
 gives the relation
\begin{eqnarray}
\bar{\alpha}^a_0 - \alpha^a_0 = \sqrt{2} w^a \ .
\label{winding}
\end{eqnarray}
Note that $w^a$ is an integer.

Using Eqs. (\ref{momentum}) and (\ref{winding}), we can get the zero modes as
\begin{eqnarray}
  \alpha^A_0 &=& \frac{1}{\sqrt{2}} G^{AB}
  \left[ p_B - (B_{BC} + G_{BC} ) w^C \right] \\
   \bar{\alpha}^A_0 &=& \frac{1}{\sqrt{2}} G^{AB}
  \left[ p_B - (B_{BC} - G_{BC} ) w^C \right] 
\end{eqnarray}
The Virasoro operators are written by 
\begin{eqnarray}
\bar{L}_0=\frac{1}{2}\bar{\alpha}^A_0\bar{\alpha}^A_0+\bar{N} \qquad
L_0=\frac{1}{2}\alpha^A_0\alpha^A_0+N
\end{eqnarray}
where $N$ and $\bar{N}$ represent the oscillators coming from Eqs. (\ref{left}) and 
(\ref{right}). We also have the level matching condition 
$L_0 -\bar{L}_0 =0$ which reads
\begin{eqnarray}
  N- \bar{N} =  p_a w^a  \ .
  \label{level}
\end{eqnarray}
It is also easy to write down the mass spectrum of a string as 
\begin{eqnarray}
\hspace{-1cm}
M^2 &=& -p^\mu p_\mu \nonumber\\
&=&  (\alpha^a_0 \alpha^a_0 
         + \bar{\alpha}^a_0 \bar{\alpha}^a_0 )
         + 2 (N + \bar{N} -2 ) \nonumber\\
&=& p_a \gamma^{ab} p_b -2 p_a \gamma^{ab} B_{bc} w^c \nonumber\\
	&&\hspace{-5mm}
	\quad+ w^a (\gamma_{ad} - B_{ab} \gamma^{bc} B_{cd} ) w^d
	+ 2 \left( N + \bar{N} -2 \right)
	\label{spectrum}
\end{eqnarray}

Let us define 
\begin{eqnarray}
   Z = \left(
   \begin{array}{c}
   p_a \\
   w^b
   \end{array}
   \right)
\end{eqnarray}
then the mass spectrum (\ref{spectrum}) and the level matching condition 
(\ref{level}) can be written as
\begin{eqnarray}
 && M^2 (Q) =  Z^T Q Z +  2 ( N + \bar{N} -2 ) \ ,
 \label{spectrum2} \\
 && N- \bar{N} = \frac{1}{2} Z^T \eta Z \ .
\end{eqnarray}
One can see the mass spectrum and the level matching condition
are invariant under $O(6,6, {\bf Z})$ 
transformation 
\begin{eqnarray}
Q\rightarrow\tilde{Q} = \Lambda^T Q \Lambda \ ,\quad
Z\rightarrow\tilde{Z} = \Lambda^{-1} Z
\end{eqnarray}
where $\Lambda \in O(6,6, {\bf Z})$ is the integer valued 
$12 \times 12$ matrix satisfying $\Lambda^T \eta \Lambda =\eta$.
As $Z$ is an integer valued
vector, $O(6,6,{\bf R})$ symmetry does not exist.

The basic assumption made in string gas cosmology is the adiabaticity
in the following sense. We assume the matter action can be represented 
by the action of the modes of the string theory
 on the torus with constant $G_{AB}$ and $B_{AB}$ 
 replaced by functions of 4-dimensional
 coordinates as $G_{AB} (x^\mu )$ and $B_{AB} (x^\mu )$. 
 The resulting action will be invariant
 under the T-duality transformation. 
 Let us imagine a gas of string consists of modes which become massless
 at the self-dual point. This is legitimate at low energy.
The energy of a string can be written as 
$ \sqrt{g^{ij} p_i p_j + M^2 (Q)} $ 
where $p_i$ is the 3-dimensional external momentum.
Hence, the energy density of the gas becomes
\begin{eqnarray}
 \rho = \frac{\mu_4}{\sqrt{g_s}}
 \sqrt{ g^{ij} p_i p_j + M^2 (Q) } \ ,
\end{eqnarray}
where $\mu_4$ is the comoving number density of a string gas in 4-dimensions
 and $g_s$ denotes the determinant of the spatial part of the 
 4-dimensional metric.
Finally, the action for the string gas is given by
\begin{eqnarray}
  S_{gas} = - \int d^{4} x \sqrt{-g} \rho \ .
\end{eqnarray}
It is not apparent this action leads the stability of moduli
 as expected. 
To grasp the feeling, we shall present the numerical results
in the next section.


\section{Evidence of stability of Dilaton}

We consider the simple situation, $B_{ab}=0$ and 
\begin{eqnarray}
  ds^2 = -dt^2 + e^{2\lambda(t)} \delta_{ij} dx^i dx^j
     + e^{2\nu (t) } \delta_{ab} dy^a dy^b  \ ,
\end{eqnarray}
where $\lambda$ and $\nu$ represents the scale factor of the 4-dimensional
universe and the radion, respectively.
This system has the symmetry under the T-duality transformation
\begin{eqnarray}
  \nu \rightarrow -\nu \ ,
\end{eqnarray}
which guarantees the stability of the radion of the internal space.
To confirm this,
we have solved the following equations numerically:
\begin{eqnarray}
&&  \ddot{\lambda} + 3\dot{\lambda}^2 
  + 6 \dot{\nu} \dot{\lambda} - 2 \dot{\lambda} \dot{\phi}
  =  \kappa^2 e^{2\phi} p_\lambda   \\
&&   \ddot{\nu} + 3\dot{\lambda} \dot{\nu} 
  + 6 \dot{\nu}^2 - 2 \dot{\nu} \dot{\phi}
  =  \kappa^2 e^{2\phi} p_\nu \\
&&  \ddot{\phi} + 3\dot{\lambda} \dot{\phi} 
  + 6 \dot{\nu} \dot{\phi} - 2 \dot{\phi}^2
  =  \frac{\kappa^2 }{2} e^{2\phi} T
  \label{dilaton}  
\end{eqnarray}
where a dot denotes a t-derivative in this section
and $T=-\rho+3p_\lambda+6p_\nu$ is 
the trace part of the energy momentum tensor 
of the string gas. 
We consider the string gas consists of the massless modes 
at the self dual point $\nu =0$ with
$N=1 , \bar{N}=0, p_a p_a =1 , w^a w^a =1 , p_a w^a =1$ which can be read
off from Eqs. (\ref{level}) and (\ref{spectrum}). 
Thus, the pressure $p_\lambda , p_\nu $ due to the string gas
 are given by
\begin{eqnarray}
p_\lambda&=&\frac{\mu_4}{e^{3\lambda}e^{6\nu}}
	\frac{p^ip_i/3}{\sqrt{
	e^{-2\lambda}p^ip_i
	+\left(e^{-\nu}- e^\nu \right)^2}}\\
p_\nu&=&\frac{\mu_4}{e^{3\lambda}e^{6\nu}}
	\frac{e^{-2\nu}- e^{2\nu}}{\sqrt{
	e^{-2\lambda}p^ip_i
	+\left( e^{-\nu}- e^\nu\right)^2}} \ .
\end{eqnarray}
The hamiltonian constraint
\begin{eqnarray}
&&\hspace{-1cm}
6 \dot{\lambda}^2 + 30 \dot{\nu}^2  
	- 12 \dot{\lambda} \dot{\phi} + 36 \dot{\nu} \dot{\lambda}
	- 24 \dot{\nu} \dot{\phi} + 4\dot{\phi}^2 
	= e^{2\phi} \rho  
\end{eqnarray}
is used to set the initial conditions. Here, the energy density of the
massless string is given by
\begin{eqnarray}
  \rho = \frac{\mu_4}{e^{3\lambda} e^{6\nu} } 
         \sqrt{e^{-2\lambda}p^i p_i + \left( e^{-\nu} - e^\nu \right)^2 } \ .
\end{eqnarray}

The results seen in Fig.1 and Fig.2 
shows the stability of the radion and the dilaton.
\begin{figure}[h]
\includegraphics[height=4cm, width=7cm]{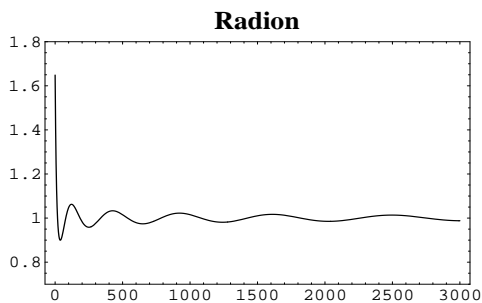}
\caption{ The radion is stabilized at the self-dual radius
independent from the initial conditions.   }
\label{fig:1}
\end{figure}
\begin{figure}[h]
\includegraphics[height=4cm, width=7cm]{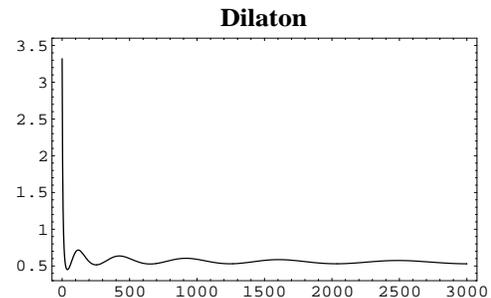}
\caption{ The dilaton is stabilized at the value which depends on the initial
conditions.  }
\label{fig:2}
\end{figure}
It is useful to rewrite the equation of motion for the dilaton (\ref{dilaton}) 
as
\begin{eqnarray}
 \frac{d^2}{dt^2} (e^{-2\phi} ) 
 + 3\dot{\lambda} \frac{d}{dt} (e^{-2\phi} )
 + 6\dot{\nu} \frac{d}{dt} (e^{-2\phi} ) + \kappa^2 T 
   = 0
\end{eqnarray}
When the radion is stabilized due to T-duality, 
the last two terms vanish because the 
energy-momentum tensor of the string gas is traceless $T=0$ 
at the self-dual radius $\nu =0$. 
In that case, the dilaton is stabilized due to the Hubble damping.

\section{T-duality invariant effective action}

In order to understand the result of numerical calculation,
we shall construct the T-duality invariant 4-dimensional effective action.
In the previous effective action approach~\cite{Battefeld:2004xw}, 
as the shifted dilaton is not used,  
the procedure of the dimensional reduction is complicated.
Moreover, T-duality symmetry is not manifest.
Hence, we use the shifted dilaton and keep the
T-duality symmetry manifest to circumvent these problems.

In order to see the stability of the moduli, we need to move on
to the Einstein frame.
Performing the conformal transformation, 
$g_{\mu\nu} = e^{2\bar{\phi}} \bar{g}_{\mu\nu}$, to the action (7),
 we obtain
\begin{eqnarray}
  S= \frac{V_6}{2\kappa^2} \int d^{4} x 
        \sqrt{-\bar{g}}
 \left[ \bar{R} -  2 (\partial \bar{\phi})^2 
 + \frac{1}{8} {\rm Tr} \partial^\mu {Q} \partial_\mu Q^{-1}   \right] \ .
\end{eqnarray}
Thus, the shifted dilaton $\bar{\phi}$ and the matrix $Q$
 are separated. 
The action for the string gas is transformed to
\begin{eqnarray}
  S_{gas} = - \int d^{4} x \sqrt{-\bar{g}} \bar{\rho} 
\end{eqnarray}
which can be interpreted as the effective potential:
\begin{eqnarray}
 \bar{\rho} &=& \frac{\mu_4}{\sqrt{\bar{g}_s}}
 \sqrt{ \bar{g}^{ij} p_i p_j 
 + e^{2\bar{\phi}} M^2 (Q) } \nonumber\\
 &\equiv& V_{\rm eff} (\bar{g}_{ij} , \bar{\phi} , Q ) \ .
 \label{potential}
\end{eqnarray}
Notice that the effective potential $V_{\rm eff}$ depends on the shifted dilaton
only through  $e^{2\bar{\phi}} $.

Suppose the moduli $Q$ are stabilized at the self-dual point.
As the mass of a string $M^2 (Q)$ vanishes at the self-dual point
by assumption, the potential of the shifted dilaton disappears.
As there exists no potential, the hubble expansion prevents the shifted
dilaton from running along the flat direction. 
Thus, the shifted dilaton  is marginally stable.
Then, the question is if the moduli $Q$ are really stabilized or not.
To investigate this issue, we need to specify the concrete
compactification model. This is discussed in the next section.

\section{Stability of $T_2 \otimes T_2 \otimes T_2$ Compactification}

We shall consider a torus compactification.
The similar but less general problem is analyzed using a different
method in \cite{Brandenberger:2005bd}
The shape moduli of the torus are completely characterized by the complex
number $\tau = \xi + i \eta $. 
The metric of the torus with unit volume is described by the metric
\begin{eqnarray}
  ds_{\rm torus}^2 = \frac{1}{{\rm Im} \tau } 
  \Big| dy^1 + \tau dy^2 \Big|^2 \ ,
\end{eqnarray}
where the periodic boundary conditions $y^1 \sim y^1 +1 , y^2 \sim y^2 +1$
 are assumed and  $|\cdots |$ denotes the absolute value of the complex number. 
\begin{figure}[h]
\includegraphics[height=5cm, width=7cm]{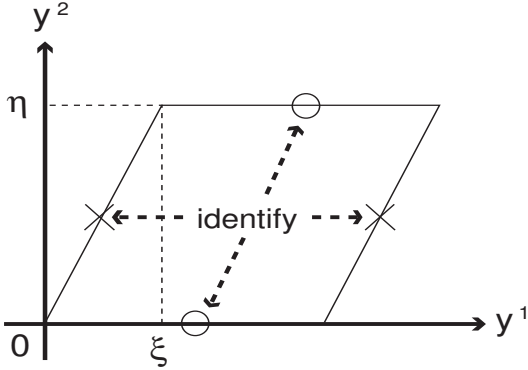}
\caption{ The shape of the torus is specified by $\xi$ and $\eta$. 
Both ends are understood to be identified.  }
\label{fig:1}
\end{figure}
The internal space we are considering is the direct product
 of the torus, $T_2 \otimes T_2 \otimes T_2 $.
 Hence, the 10-dimensional metric to be consider is
\begin{eqnarray}
\hspace{-1cm}
  ds^2 &=& \bar{g}_{\mu\nu} dx^\mu dx^\nu  
   + \sum_{a=1}^3  \frac{b_a^2}{{\rm Im} \tau_a } 
  \Big| dy^{2a-1} + \tau_a dy^{2a} \Big|^2  \ ,
\end{eqnarray}
where we have defined three scale factor $b_a $ 
 and the modulus $\tau_a$ for each torus with coordinates $y^{2a-1} ,y^{2a}$.
 We would like to analyze the stability of
$T_2 \otimes T_2 \otimes T_2 $ compacification. 
Fortunately, as the internal space is the direct product
of torus, it is enough to investigate the simple 6-dimensional
spacetime with one torus as the internal space.

Now, we shall take the metric  
\begin{eqnarray}
  ds^2 &=& \bar{g}_{\mu\nu} dx^\mu dx^\nu  
  \label{general}\\
       && \quad  + \frac{b^2}{\eta} \left[ (dy^1 + \xi dy^2 )^2 
         + \eta^2 (dy^2 )^2 \right] 
         \ , \nonumber
\end{eqnarray}
where $\bar{g}_{\mu\nu}$ is the 4-dimensional metric in the Einstein frame
and  $b$ represents the scale factor of the torus, i.e. volume moduli. 
The anti-symmetric tensor field in 2-dimensions has only one component
\begin{eqnarray}
 B = \left(
 \begin{array}{cc}
 0 & \beta \\
 -\beta & 0
 \end{array}
 \right)  
 \label{beta}\ .
\end{eqnarray}
We call $\beta$ the flux moduli, hereafter. 
The 4-dimensional effective action in the Einstein frame becomes
\begin{eqnarray}
  S &=& \frac{V_6}{2\kappa^2 } \int d^4 x \sqrt{-\bar{g}} 
  \left[ \bar{R} - 2 (\partial \bar{\phi} )^2 
           -2 (\partial \log b)^2 \right. \nonumber\\
 && \left. \quad - \frac{1}{2\eta^2 } \left\{ (\partial \eta )^2 
       + (\partial \xi)^2 \right\} 
       -\frac{1}{2b^4} (\partial \beta )^2 \right]
      \label{main-action}  \\
&&   - \mu_4 \int d^4 x \sqrt{-\bar{g}_{00}} \sqrt{\bar{g}^{ij} p_i p_j 
          + e^{2\bar{\phi}} M^2 (\beta , b,\eta , \xi) }  \ ,  \nonumber 
\end{eqnarray}
where the mass is given by
\begin{eqnarray}
  M^2 (\beta , b,\eta , \xi)
  &=& \frac{1}{\eta b^2} \left\{ p_1 \xi - p_2 - 
              \beta ( w^2 \xi + w^1 )\right\}^2  \nonumber\\      
  &&  + \frac{\eta}{b^2} ( p_1 - w^2 \beta )^2 
  + \frac{b^2}{\eta} (w^2 \xi + w^1 )^2 \nonumber \\
  && + \eta b^2 (w^2 )^2
  +  4 (N-1) \nonumber\\
  &&  - 2 (p_1 w^1 + p_2 w^2 )  \ .
\end{eqnarray}
Using Eq.~(\ref{beta}) and 
\begin{eqnarray}
\Gamma&=& \frac{b^2}{\eta}\left(
	\begin{array}{cc}
	1 & \xi \\
	\xi &\ \xi^2+\eta^2
	\end{array}
	\right)
\end{eqnarray}
which one can read off from the metric (\ref{general}).
We see the above action has the T-duality symmetry 
(\ref{Gamma}) and (\ref{B}) :
\begin{eqnarray}
 && \tilde{\eta} =\frac{\eta}{\eta^2 + \xi^2} \ , \quad
  \tilde{\xi} =- \frac{\xi}{\eta^2 + \xi^2} \nonumber\\
 &&  \tilde{b}^2 =\frac{b^2}{b^4 + \beta^2} \ , \quad
  \tilde{\beta} =- \frac{\beta}{b^4 + \beta^2} \ .
  \label{symmetry}
\end{eqnarray}
From Eq.~(\ref{symmetry}), it is easy to find the self-dual point 
\begin{eqnarray}
  b=1 \ ,\quad  \eta =1 \ ,\quad  \xi =0 \ ,\quad  \beta =0 \ .
\end{eqnarray}

One may expect this self-dual point is a stable minimum
 of the effective potential. In order to verify this, 
 we should examine where is the minimum of $M^2(\beta , b,\eta , \xi)$
 in the potential in the action (\ref{main-action}). 
First, let us consider the string gas consisting of modes 
$N=1 , p_1 = w^1 =1 , p_2 = w^2 =0$ which becomes  massless 
at the self-dual point. For this gas, we have 
\begin{eqnarray}
 M^2_1 = \frac{1}{\eta b^2} (\xi -\beta)^2 
 + \frac{\eta}{b^2} + \frac{b^2}{\eta} -2
\end{eqnarray}
In this case, there exists flat directions $b^2 = \eta \ ,\xi =\beta$
 in contrast to the naive expectation.
However, we only considered one kind of string gas which winds around
one specific cycle. Apparently, we have the other cycle for the torus.
Hence, we consider another string gas consisting of modes 
$N=1 , p_1 = w^1 = 0 , p_2 = w^2 =1 $ which becomes massless at the
self-dual point.  In this case, we obtain
\begin{eqnarray}
 M^2_2 = \frac{1}{\eta b^2} (1+\beta\xi)^2 
     + \frac{\eta \beta^2}{b^2} + \frac{b^2 \xi^2 }{\eta} 
     +\eta b^2 -2 \ .
\end{eqnarray}
Now, we also have flat directions  $\beta^2 = \frac{b^4\xi^2 }{\eta^2}
  \ , 1+\beta \xi=\eta b^2 $.
We find these two flat directions  intersect at the self-dual point 
$b=1 \ , \eta =1 \ , \xi =0, \beta =0$.   
Hence, by taking into account both type of string gas,   
 the self-dual point would be stable minimum.
The stability can be explicitly verified 
by expanding the potential around this extrema as 
\begin{eqnarray}
  V = \mu_4 \sqrt{\frac{\bar{g}^{ij} p_i p_j}{g_s}}
  + \frac{1}{2}\frac{\mu_4 e^{2\bar{\phi}}}{\sqrt{\bar{g}^{ij} p_i p_j}}
  M^2 (\beta, b , \eta , \xi) 
\end{eqnarray}
where we have used the fact that $M^2 \approx 0$ near the self-dual point.
Let us linearize the scale factor $b$ and the modulus $\eta$
as $b= 1+ \delta b$ and $ \eta = 1 + \delta \eta$. Other variables 
$\xi $ and $\beta$ are already linear because the background values
of these variables are zero. Hence, we have 
\begin{eqnarray}
  \delta M^2_1 = (\xi -\beta)^2 + (\delta \eta -2 \delta b)^2
\end{eqnarray}
and
\begin{eqnarray}
  \delta M^2_2 = (\xi +\beta)^2 + (\delta \eta +2 \delta b)^2 \ .
\end{eqnarray}
Here, we can see each potential has flat directions.
However, by adding up both contributions, we get
\begin{eqnarray}
  \delta M^2 = \delta M^2_1 + \delta M^2_2 
  = 2\xi^2 +2\beta^2 + 2\delta \eta^2 +8 \delta b^2
\end{eqnarray}
where flat directions disappear.  
Thus, we have proved the stability of all of the moduli of the torus.
Even if we add other massless modes, the result does not change.
The dilaton is also stabilized due to the reason explained
in the previous section. 
This concludes the stability of
$T_2 \otimes T_2 \otimes T_2 $ compacification as we expected.

\section{Conclusion}

 We have analyzed the stability  of
 $T_2 \otimes T_2 \otimes T_2$ compactification
 in the context of massless string gas cosmology. We emphasized
 the importance of the T-duality and massless modes in a string.
 We have first performed the numerical calculations and then shown the
 stability of the dilaton. To understand this numerical result
 we have constructed the 4-dimensional effective action by taking 
 into account the T-duality. It turned out that the dilaton is
 marginally stable.
 We performed the stability analysis of
 the volume moduli, the shape moduli and the flux moduli. 
 We have found that all of these 
 moduli are stabilized at the self dual point in the moduli space.

Of course, what we have shown is the stability of moduli during the string 
gas dominated stage. 
After the string gas dominated stage, the ordinally matter come to dominate
the universe. Then, the dilaton will start to run.
Therefore, we need to find a mechanism to stabilize the dilaton 
in these periods. One possibility is the non-perturbative string
correction 
\begin{eqnarray}
S &=& \frac{1}{2\kappa^2 }\int d^{10} x \sqrt{-G}  
   \left[ B_1 (\phi)R + 4 B_2 (\phi) (\partial \phi )^2  \right. \nonumber \\
  && \left. \qquad \qquad - \frac{1}{12}B_3(\phi)
    H^2 \right]  \ , 
\end{eqnarray}
where we have taken into accounts the loop corrections
\begin{eqnarray}
  B_i (\phi ) = e^{-2\phi} + c_i e^{2\phi} + d_i e^{4\phi} + \cdots \ .
\end{eqnarray}
There may be a possibility to stabilize all of the moduli 
in the whole history of the universe 
in this context~\cite{Damour:1994ya,Damour:1994zq}.
This possibility deserve further investigations.
 It is also interesting to investigate the possibility to combine
 the string gas approach with other ones~\cite{Kachru,Berndsen:2005qq}.
 
 More importantly, we need a mechanism to explain the present large scale
 structure of the universe in the context of the string gas 
 scenario. 
 In the string gas model, it is difficult to realize the 
 inflationary scenario. We might have to seek a completely different
 one from the inflationary generation mechanism of primordial 
 fluctuations~\cite{Battefeld:2005wv}.

\acknowledgments 
We would like to thank Robert Brandenberger for useful discussions
and suggestions. 
This paper was initiated by discussions during the conference 
``The Next Chapter in Einstein's Legacy'' at the Yukawa Institute, Kyoto, 
and the subsequent workshop. 
  This work was supported by the
  Grant-in-Aid for the 21st Century COE "Center for Diversity and
  Universality in Physics" from the Ministry of Education, Culture,
  Sports, Science and Technology (MEXT) of Japan.  This work was also
  supported in part by Grant-in-Aid for Scientific Research Fund of
  the Ministry of Education, Science and Culture of Japan No. 155476
  (SK), No.14540258  and No.17340075 (JS).

%

\end{document}